\documentclass[a4paper,12pt]{article}
\pdfoutput=1

\usepackage{amsmath, amssymb, authblk, a4wide, bm, cancel, color, graphicx, subfig, enumerate, appendix, youngtab, cite, hyperref, bbold, verbatim, slashed}

\newcommand{\pdiff}[2]{\frac{\partial#1}{\partial#2}}

\newcommand{\dimintlim}[4]{\int_{#3}^{#4}\mathrm{d}^{#1}#2\,}
\newcommand{\nbrack}[1]{\left(#1\right)}
\newcommand{\sbrack}[1]{\left[#1\right]}
\newcommand{\expect}[1]{\langle#1\rangle}
\newcommand{\norm}[1]{\left|#1\right|}
\newcommand{\pfrac}[2]{\left(\frac{#1}{#2}\right)}

\def\fund{\tiny\Yvcentermath1\yng(1)}

\def\asym{\tiny\Yvcentermath1\yng(1,1)}
\def\widerow{\rule{0pt}{2.5ex}\rule[-1.5ex]{0pt}{0pt}}
\def\be{\begin{equation}}
\def\ee{\end{equation}}
\def\ba{\begin{eqnarray}}
\def\ea{\end{eqnarray}}

\def\cL{{\cal L}}

\def\p0{{\phantom{+}0}}

\def\uno{\mbox{1 \kern-.59em {\rm l}}}

\numberwithin{equation}{section}

\begin{document}

\title{
\rightline{\sc\small UMN-TH-3313/13}
\vspace{2cm}
\Large{\textbf{UV descriptions of composite Higgs models without elementary scalars}}}
\author[1]{\small{\bf James Barnard}\thanks{\texttt{james.barnard@unimelb.edu.au}}}
\author[2,1]{\small{\bf Tony Gherghetta}\thanks{\texttt{tgher@umn.edu}}}
\author[1]{\small{\bf Tirtha Sankar Ray}\thanks{\texttt{tirtha.sankar@unimelb.edu.au}}}
\affil[1]{ARC Centre of Excellence for Particle Physics at the Terascale, School of Physics, 
The University of Melbourne, Victoria 3010, Australia }
\affil[2]{School of Physics and Astronomy, University of Minnesota, Minneapolis, MN 55455, USA}
\date{}
\maketitle

\begin{abstract}
\baselineskip=15pt
\noindent
We consider four-dimensional UV descriptions of composite Higgs models without elementary scalars, in which four-fermion interactions are introduced to an underlying gauge theory like in the gauged NJL model.  When the anomalous dimension of the fermion bilinear is large, these interactions drive the spontaneous global symmetry breaking in the model, with the Higgs identified as a Nambu-Goldstone boson. The UV descriptions support composite top partner operators, also with large anomalous dimensions, thereby providing an explicit realisation of the idea of partial compositeness.  In particular, the composite $SO(6)/SO(5)$ model can be described by an $Sp$ gauge theory with four flavours of fermion, together with a vector-like pair of fermions transforming in the antisymmetric representation and charged under $SU(3)$ colour.  These fermions confine to produce both the Higgs and top partner bound states.  Our methods can also be applied to different coset groups, suggesting that four-fermion operators can describe the underlying UV dynamics of other composite Higgs models.
\end{abstract}

\newpage
\section{Introduction}

The recently discovered 126 GeV Higgs boson confirms that the Higgs mechanism is responsible for electroweak symmetry breaking in the Standard Model.  However, the nature of the Higgs boson remains a mystery.  One possibility is that the Higgs is an elementary particle, similar to the fermions and gauge bosons in the Standard Model, which leads to scenarios that may or may not involve supersymmetry.  Alternatively, it remains logically possible that the Higgs is a composite state of some underlying strong dynamics that appears beyond the TeV scale.  Although this is similar to QCD, it would require that the strong dynamics produces a light, CP-even scalar bound state that is much lighter than the scale of the strong dynamics.  Consequently, a mechanism is needed to suppress quantum corrections to the Higgs mass, otherwise there is a little hierarchy between the electroweak scale and the scale of the strong dynamics.

One way to protect the Higgs mass from quadratic divergences is to assume that it is a composite Nambu-Goldstone boson (NGB) state emerging from the strongly coupled sector~\cite{Kaplan:1983fs, Dugan:1984hq}.  The strong sector can spontaneously break a global symmetry, and the Higgs degrees of freedom are identified with the massless NGBs, much like the pions in QCD\@.  When the Higgs sector is coupled to the Standard Model, the global symmetry is explicitly broken by the gauge and Yukawa interactions, giving rise to electroweak symmetry breaking and a Higgs mass.  Furthermore, it is necessary that the top quark couples relatively strongly to the Higgs sector, so as to efficiently communicate the explicit global symmetry breaking.  This is typically achieved by a marginally relevant bilinear mixing term between the elementary top quark and a composite top partner operator.\footnote{A Higgs mass of 126 GeV usually requires light top partner resonances~\cite{Matsedonskyi:2012ym, Marzocca:2012zn, Pomarol:2012qf, Panico:2012uw, Pappadopulo:2013vca, Barnard:2013hka}, although alternative approaches are possible \cite{Cheng:2013qwa}.}  The same mixing can also, potentially, explain the origin of the fermion mass hierarchy, as the composite Higgs will couple much more strongly to composite fermions than elementary fermions~\cite{Kaplan:1991dc, Gherghetta:2000qt}.  Hence the top quark is a mostly composite state while the other quarks and leptons are, to varying degrees, mostly elementary.

The ever present problem with this idea is the presence of a strong sector, which makes a full understanding of the relevant physics somewhat elusive.  A popular approach is to not specify the details of the strongly coupled UV physics and instead concentrate on a low-energy, effective theory based on a chiral Lagrangian~\cite{Giudice:2007fh, Barbieri:2007bh, Anastasiou:2009rv, Gripaios:2009pe, Mrazek:2011iu, Panico:2011pw, DeCurtis:2011yx, Contino:2011np, Alonso:2012px, Brivio:2013pma}.  Chiral Lagrangians depend only on the spontaneous global symmetry breaking pattern and the choice of fermion representations, yet still provide a semi-quantitive description of the low-energy physics.  However, while there are many choices for phenomenologically viable chiral Lagrangians, there is no guarantee that there exists a suitable UV description that they can emerge from.  Specifically, the main purpose for assuming a composite Higgs is to protect the Higgs mass from the large corrections it receives due to its scalar nature.  If the UV description behind the chosen chiral Lagrangian necessarily contains light scalars, the problem is simply moved to a different sector.

Two main approaches have been suggested to deal with the UV physics of these models.  In supersymmetric gauge theories strong coupling can be much better understood by using ideas such as Seiberg duality~\cite{Seiberg:1994pq, Intriligator:1995au} (itself closely related to the notion of chiral Lagrangians~\cite{Abel:2012un}) and there have been realisations of composite Higgs models with a well defined, supersymmetric UV description~\cite{Caracciolo:2012je}.  The down side is that supersymmetry necessarily introduces many elementary scalars to the model, although it does protect them from quadratic divergences.  Alternatively, one can use the AdS/CFT correspondence to recast a model of electroweak symmetry breaking in a strongly coupled, four-dimensional CFT to a weakly coupled model in a slice of $AdS_5$~\cite{Contino:2003ve, Agashe:2004rs}.  However, a true UV description in five dimensions relies on knowing the complete (string) theory of gravity.

Both approaches rely on adding a large amount of extra structure to the model that makes the original solution to the hierarchy problem -- that the Higgs is a NGB -- redundant, as both require a separate method for stabilising scalar masses.  The questions we therefore ask here are, given a low-energy effective description based on a chiral Lagrangian, is it possible to find a well defined UV description, containing only fermions and gauge fields, such that the Higgs mass is protected by its NGB status only?  And, if so, what sort of properties must the UV description possess?  Earlier work suggests such models are indeed possible~\cite{Miransky:1988xi, Galloway:2010bp} but, as yet, only for specific chiral Lagrangians and with no notion of partially composite top quarks.\footnote{Ref.~\cite{Miransky:1988xi} forms the Higgs from a top condensate, and the global symmetry is broken within the strongly coupled sector in Ref.~\cite{Galloway:2010bp}.}

A convenient way to address these questions is to introduce four-fermion operators to the model, which are well known to drive spontaneous symmetry breaking in the Nambu-Jona Lasinio (NJL) model~\cite{Nambu:1961tp}. We consider the $SO(6)/SO(5)$ model~\cite{Batra:2007iz,Gripaios:2009pe}, containing a Higgs doublet and a singlet, by realising the $SO(6)$ global symmetry in an $Sp(2N_c)$ gauge theory with four flavours of Weyl fermion. After adding four-fermion operators to this gauge theory, we study the resulting gauged NJL model and show that the $SO(6)$ global symmetry can be spontaneously broken to $SO(5)$.  In particular, the symmetry breaking is induced mainly by the four-fermion operators when the anomalous dimension of the constituent fermion bilinear is large. The same large anomalous dimension can then be used to obtain large anomalous dimensions for the top partners, which helps to explain the large top mass in composite Higgs models.  A similar mechanism may also work for other composite Higgs models, but the analysis in these cases is more complicated.

The outline of this paper is as follows.  In section \ref{sec:65} we study the $SO(6)/SO(5)$ model, and show in section~\ref{sec:gNJL} that this symmetry breaking pattern follows from a gauged NJL model with an $Sp(2 N_c)$ gauge symmetry and four flavours of Weyl fermion.  In particular, in section \ref{sec:LA} we show that there are two ways in which the global symmetry can be broken, with the fermion bilinear having either a large or small anomalous dimension.  The fermion bilinear is used in section \ref{sec:65c} to obtain a composite top partner by introducing an additional pair of coloured fermions.  In section \ref{sec:ADtop} the mixing between this top partner and the elementary top is shown to be almost relevant when the anomalous dimension of the fermion bilinear is large, leading to a sufficiently large top quark mass and successful electroweak symmetry breaking.  In section \ref{sec:UVobs} we briefly outline UV observables that can be computed in our UV description.  Other composite Higgs models are considered in section \ref{sec:other}, where we discuss how to address issues in the minimal composite Higgs model $SO(5)/SO(4)$ as well as more general models.  Our summary is given in section \ref{sec:sum}.  Details of minimising the potential in the $SO(6)$ model are given in the Appendix.

\section{The $SO(6)/SO(5)$ model\label{sec:65}}

A simple example, which illustrates many important features of potential UV descriptions, is the next to minimal composite Higgs model with spontaneous symmetry breaking pattern $SO(6)\to SO(5)$.  The Higgs sector of this model consists of a SM Higgs doublet and a CP-odd singlet scalar, which has interesting phenomenological consequences~\cite{Gripaios:2009pe}.  To find a UV description of this model we will instead use the equivalent symmetry breaking pattern $SU(4)\to Sp(4)$.

This symmetry breaking can occur in an $Sp(2N_c)$ gauge theory with four flavours of two component Weyl fermion, $\psi$, the transformation properties of which are summarised in table \ref{tab:65}.\footnote{We will work with two component Weyl fermions throughout this paper.}  The $SU(4)$ global symmetry appears as an exchange symmetry acting on the fermions, and is automatically present in the absence of any interaction terms in the Lagrangian (the $U(1)$ that would complete the symmetry to $U(4)$ is anomalous).  Since the gauge group is symplectic the model does not suffer from a chiral anomaly, and an even number of flavours (or, more specifically, the sum of the Dynkin indices of all chiral fermions is even) ensures the absence of a Witten anomaly.

The gauge sector is asymptotically free for any value of $N_c$.  To see this, recall that the one loop beta function coefficient is given by
\be\label{eq:bdef}
b=\frac{11}{3}C_2({\bf adj})-\frac{2}{3}\sum_iT(r_i)N_f(r_i)
\ee
for the quadratic Casimir of representation $r$, $C_2(r)$, where
\be
T(r)=C_2(r)\frac{\dim(r)}{\dim(\bf adj)}
\ee
and where $N_f(r)$ is the number of fermions in representation $r$.  For an $Sp(2N_c)$ gauge theory $C_2({\bf adj})=2N_c+2$ and $T(\fund)=1$.  As such
\be\label{eq:b65}
b=\frac{11}{3}(2N_c+2)-\frac{2}{3}\times4=\frac{2}{3}(11N_c+7)
\ee
when $N_f=4$, which is positive for any value of of $N_c$.

\begin{table}[!t]
\begin{equation*}
\begin{array}{|c|c|c|}\hline
\widerow & Sp(2N_c) & SU(4) \\\hline
\widerow \psi & \fund & {\bf4} \\\hline\hline
\widerow M & {\bf1} & {\bf6} \\\hline
\end{array}
\end{equation*}
\caption{A model that realises the spontaneous symmetry breaking pattern $SU(4)\to Sp(4)$, where $\psi$ is a two component Weyl fermion, and $M$ is a complex scalar meson.\label{tab:65}}
\end{table}

Because the gauge theory is asymptotically free it is expected to confine, whereupon a fermion condensate will spontaneously break the global symmetry \cite{Coleman:1980mx,Peskin:1980gc}.  To formally determine whether and in what way the $SU(4)$ symmetry is broken we will add four-fermion operators to the Lagrangian, whereupon the model becomes a variant of the gauged NJL model.  Four-fermion operators will turn out to be very important when calculating the details of the spontaneous symmetry breaking.  If they are not added by hand, the operators we will consider could instead be generated by $Sp(2N_c)$ dynamics (c.f.\ instantons in QCD~\cite{Diakonov:1995e a, Schafer:1996wv}) so the model can be considered as an effective, low-energy description of the $Sp(2N_c)$ gauge theory that is used to determine the model's vacuum structure.  Or, and perhaps more usefully, the four-fermion operators can be considered as bare terms in a strongly interacting UV description, provided that the fermions have a large enough anomalous dimension to make the operators relevant (and therefore the UV description renormalisable).  We will discuss both options in more detail shortly.

\subsection{An $SU(4)$ gauged NJL model\label{sec:gNJL}}
The leading order fermion interaction Lagrangian, constructed out of gauge invariant bilinears and respecting the global symmetry, 
is
\be\label{eq:Lint}
\cL_{\rm int}=\frac{\kappa_A}{2N_c}(\psi^a\psi^b)(\bar{\psi}_a\bar{\psi}_b)+\frac{\kappa_B}{8N_c}\sbrack{\epsilon_{abcd}(\psi^a\psi^b)(\psi^c\psi^d)+\mbox{h.c.}}
\ee
where $a,b,c,d$ are $SU(4)$ indices, $\epsilon$ is the Levi-Civita symbol and $\kappa_{A,B}$ are real, dimensionful coupling constants. Fermion bilinears are defined according to
\be
(\psi^a\psi^b)\equiv\Omega_{ij}\psi_i^a\psi_j^b
\ee
where $\Omega$ is the antisymmetric, $2N_c\times 2N_c$ matrix
\be\label{eq:Omega}
\Omega=\begin{pmatrix} 0 & \uno_{N_c} \\ -\uno_{N_c} & 0 \end{pmatrix}
\ee
used to contract gauge indices in $Sp$ gauge theories.  Since $\Omega$ is an antisymmetric tensor, the bilinears are antisymmetric in the $SU(4)$ indices $a,b$.  There are other operators, such as $(\Omega_{ik}\psi_i^a\psi_j^b)(\Omega_{jl}\bar{\psi}_{ka}\bar{\psi}_{lb})$, that we could add to the Lagrangian, but these can be rewritten as vector-vector operators, such as $(\Omega_{ik}\psi_i^a\sigma^\mu\bar{\psi}_{ka})(\Omega_{jl}\psi_j^b\sigma_\mu\bar{\psi}_{lb})$, by applying Fierz identities.  Hence they describe interactions between vector mesons and play no part in the spontaneous symmetry breaking.  We will, however, return to them later when discussing top partners in this model, and those with electroweak charge are well known to be important in maintaining perturbative unitarity.

By introducing an auxiliary, antisymmetric scalar field $M$, transforming as in table \ref{tab:65}, and couplings
\begin{align}\label{eq:Laux}
\cL_{\rm aux}=-{} & \frac{\kappa_A}{2N_c}\nbrack{(\psi^a\psi^b)+\frac{2N_c}{\kappa_A+\kappa_B}M^{ab}}\nbrack{(\bar{\psi}_a\bar{\psi}_b)+\frac{2N_c}{\kappa_A+\kappa_B}M^*_{ab}}-{} \nonumber\\
& \frac{\kappa_B}{8N_c}\sbrack{\epsilon_{abcd}\nbrack{(\psi^a\psi^b)+\frac{2N_c}{\kappa_A+\kappa_B}M^{ab}}\nbrack{(\psi^c\psi^d)+\frac{2N_c}{\kappa_A+\kappa_B}M^{cd}}+\mbox{h.c.}}
\end{align}
the interaction Lagrangian can be rewritten as
\begin{align}\label{eq:Leff}
\cL_{\rm int}=-{} & \frac{1}{\kappa_A+\kappa_B}\sbrack{\nbrack{\kappa_AM^*_{ab}+\frac{1}{2}\kappa_B\epsilon_{abcd}M^{cd}}(\psi^a\psi^b)+\mbox{h.c.}}-{} \nonumber\\
& \frac{2N_c\kappa_A}{(\kappa_A+\kappa_B)^2}M^{ab}M^*_{ab}-\frac{N_c\kappa_B}{2(\kappa_A+\kappa_B)^2}\nbrack{\epsilon_{abcd}M^{ab}M^{cd}+\mbox{h.c.}}
\end{align}
with no effect on the physics.  The model now resembles massive Yukawa theory, enabling a much simpler derivation of the vacuum structure.  The auxiliary scalar $M$ has been constructed so as to describe the vacuum expectation value (VEV) of the scalar meson fermion bilinear, $\psi\psi$, as can be seen from the solution to its equation of motion
\be\label{eq:Mdef}
M^{ab}=-\frac{\kappa_A+\kappa_B}{2N_c}(\psi^a\psi^b).
\ee
The prefactor for $\psi\psi$ is actually arbitrary, but the above choice will turn out to be convenient later.

The matrix $M$ is complex and antisymmetric, so can always be rotated into the form
\be\label{eq:M4vev}
M=\begin{pmatrix}
0 & m_1 & 0 & 0 \\
-m_1 & 0 & 0 & 0 \\
0 & 0 & 0 & m_2 \\
0 & 0 & -m_2 & 0 \end{pmatrix}
\ee
by an $SU(4)$ transformation for some undetermined constants $m_{1,2}$ ($i$ times the eigenvalues of $M$) whereupon the interaction Lagrangian becomes
\begin{align}\label{eq:LSp}
\cL_{\rm int}=-{} & \frac{1}{\kappa_A+\kappa_B}\sbrack{\nbrack{\kappa_AM^*_{ab}+\frac{1}{2}\kappa_B\epsilon_{abcd}M^{cd}}(\psi^a\psi^b)+\mbox{h.c.}}-{} \nonumber\\
& \frac{4N_c\kappa_A}{(\kappa_A+\kappa_B)^2}(|m_1|^2+|m_2|^2)-\frac{4N_c\kappa_B}{(\kappa_A+\kappa_B)^2}(m_1m_2+m_1^*m_2^*).
\end{align}
When $M$ obtains a VEV (i.e.\ a fermion condensate forms) the Yukawa coupling dynamically generates a Dirac mass for the fermions.  The mass-squared eigenvalues are
\begin{align}\label{eq:barm}
|\bar{m}_1|^2 & =\frac{4|\kappa_Am_1^*+\kappa_Bm_2|^2}{(\kappa_A+\kappa_B)^2} &
|\bar{m}_2|^2 & =\frac{4|\kappa_Am_2^*+\kappa_Bm_1|^2}{(\kappa_A+\kappa_B)^2}
\end{align}
so, at one-loop order, the effective potential for the eigenvalues of $M$ is
\begin{align}
V(m)={} & \frac{N_c}{\kappa_A^2-\kappa_B^2}\sbrack{\kappa_A(|\bar{m}_1|^2+|\bar{m}_2|^2)-\kappa_B(\bar{m}_1\bar{m}_2+\bar{m}_1^*\bar{m}_2^*)}-{} \nonumber\\
& \frac{N_c}{8\pi^2}\sum_{i=1}^2\sbrack{\Lambda^2|\bar{m}_i|^2+|\bar{m}_i|^4\ln\pfrac{|\bar{m}_i|^2}{\Lambda^2+|\bar{m}_i|^2}+\Lambda^4\ln\pfrac{\Lambda^2+|\bar{m}_i|^2}{\Lambda^2}}.
\end{align}
The scale $\Lambda$ is a UV cutoff scale used to regularise the loop integrals.  We will treat it as an arbitrary scale for now, but will discuss its interpretation in more detail shortly.\footnote{Note also that the cutoff can be introduced in a gauge-invariant way that preserves the quadratically divergent terms by using dispersion relations~\cite{Gherghetta:1994cr}}  Returning to the problem at hand, the $\bar{m}_1\bar{m}_2$ term in the potential is the only term dependent on the phases of the eigenvalues.  It is minimised when $\bar{m}_1\bar{m}_2=|\bar{m}_1||\bar{m}_2|$ if $\kappa_B/(\kappa_A^2-\kappa_B^2)>0$, or when $\bar{m}_1\bar{m}_2=-|\bar{m}_1||\bar{m}_2|$ if $\kappa_B/(\kappa_A^2-\kappa_B^2)<0$.  We therefore take both $\bar{m}$'s to be real and positive to find the final form for the potential
\begin{align}\label{eq:VSp1}
V(m)={} & \frac{N_c\kappa_A}{\kappa_A^2-\kappa_B^2}(\bar{m}_1^2+\bar{m}_2^2)-\norm{\frac{2N_c\kappa_B}{\kappa_A^2-\kappa_B^2}}\bar{m}_1\bar{m}_2-{} \nonumber\\
& \frac{N_c}{8\pi^2}\sum_{i=1}^2\sbrack{\Lambda^2\bar{m}_i^2+\bar{m}_i^4\ln\pfrac{\bar{m}_i^2}{\Lambda^2+\bar{m}_i^2}+\Lambda^4\ln\pfrac{\Lambda^2+\bar{m}_i^2}{\Lambda^2}}.
\end{align}

The above potential always has a stationary point at the origin, $\bar{m}_1=\bar{m}_2=0$ (which also means $m_1=m_2=0$) where the $SU(4)$ symmetry is not broken.  There can be a second stationary point at $\bar{m}_1=\bar{m}_2=\bar{m}$ (see appendix \ref{app:pot}) and
\be\label{eq:mmin}
1-\frac{\bar{m}^2}{\Lambda^2}\ln\pfrac{\Lambda^2+\bar{m}^2}{\bar{m}^2}=\frac{4\pi^2}{\Lambda^2}\nbrack{\frac{\kappa_A}{\kappa_A^2-\kappa_B^2}-\norm{\frac{\kappa_B}{\kappa_A^2-\kappa_B^2}}}\equiv\frac{1}{\xi}
\ee
where we have defined a new, dimensionless coupling $\xi$.  Returning to the $(m_1,m_2)$ basis via eq.~\eqref{eq:barm}, this stationary point turns out to be at
\be
m_1=m_2=\frac{\bar{m}}{2}
\ee
and it exists if and only if $\xi>1$, i.e.\ the four-fermion coupling is strong.  It is then always the global minimum of the potential.  Hence $\xi=1$ corresponds to a critical point separating phases of broken and unbroken global symmetry.  Because $m_1=m_2$ in eq.~\eqref{eq:M4vev}, the VEV of $M$ is of the correct form to break the $SU(4)$ global symmetry down to $Sp(4)$ and the model is a good candidate for producing the Higgs as an NGB\@.

\begin{figure}[!t]
\begin{center}
\includegraphics[width=0.45\textwidth]{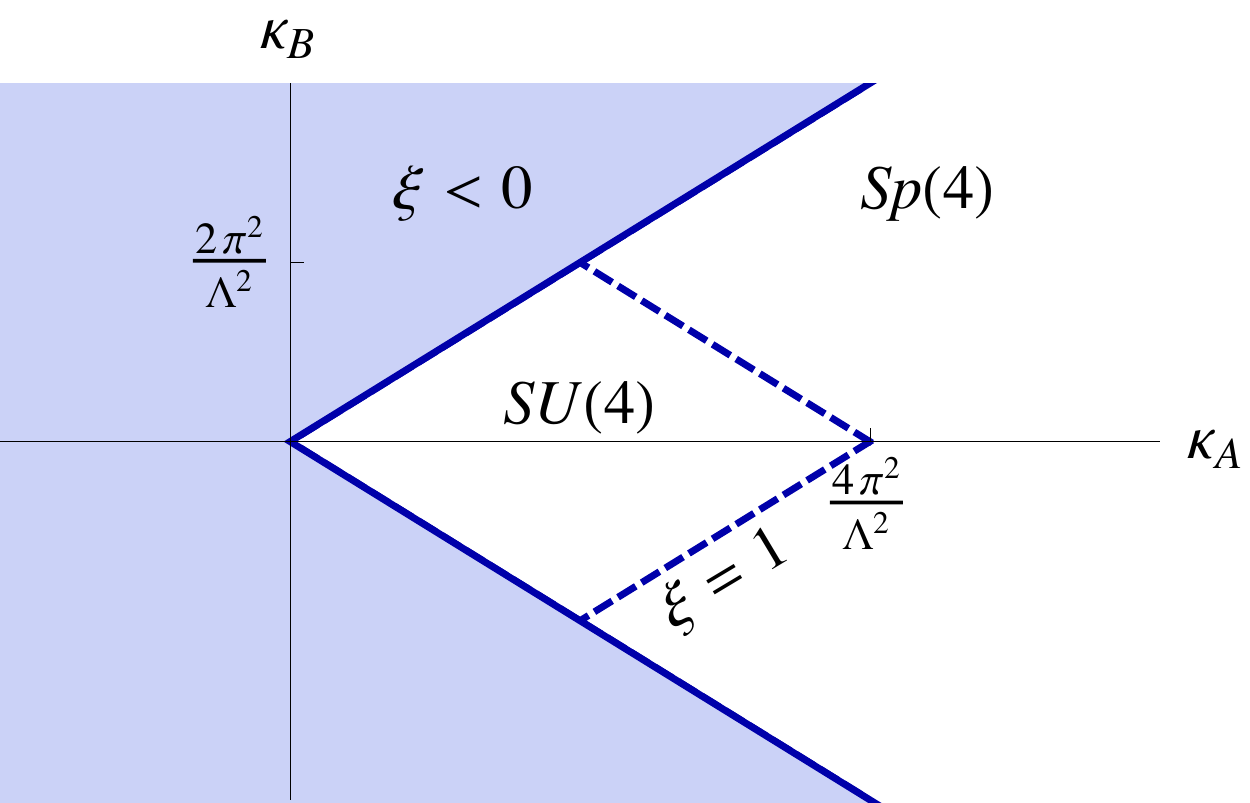}
\end{center}
\caption{The phase diagram in the $(\kappa_A,\kappa_B)$ plane.  In the shaded region $\xi<0$ and the details of the vacuum are incalculable.  In the region inside the central box $0<\xi<1$ and the global symmetry is unbroken.  In the region outside the central box $\xi>1$ and the $SU(4)$ global symmetry is broken to $Sp(4)$.\label{fig:kappa}}
\end{figure}

The potential is again unstable around the origin, this time at tree level, if $\xi<0$ (i.e.\ $\kappa_A<0$ or $\kappa_A^2<\kappa_B^2$) and we again expect some kind of spontaneous symmetry breaking.  However, in this case we cannot follow the steps above to determine the VEV of the fermion condensate.  Eq.~\eqref{eq:mmin} would yield a solution with $\bar{m}>\Lambda$, thereby invalidating our expression for the one-loop effective potential and rendering the details of the minimum, and the symmetry breaking pattern, incalculable.  Putting everything together, we therefore find a phase diagram in the $(\kappa_A,\kappa_B)$ plane as illustrated in figure \ref{fig:kappa}.

\subsubsection{Large anomalous dimensions\label{sec:LA}}

The scale $\Lambda$ was introduced as a UV cutoff when calculating the one-loop effective potential, and persists in the expression \eqref{eq:mmin} for the symmetry breaking VEV\@.  It has two interpretations.  It could be considered as a physical cutoff for the effective Yukawa theory, i.e.\ the scale at which the four-fermion operators are generated.  As the four-fermion operators preserve the full $SU(4)$ exchange symmetry acting on $\psi$, they have no effect on the symmetries of the initial gauge theory.  Hence it is possible to generate them via the $Sp(2N_c)$ dynamics alone, whereupon $\Lambda$ can be identified with the strong coupling scale of the $Sp(2N_c)$ gauge group.  For generic values of the coupling $\xi$ one then expects $\bar{m}\sim\Lambda$, due to eq.~\eqref{eq:mmin}.  Smaller values for the VEV can only be achieved if $\xi$ is tuned to be very close to one.

Instead of considering $\Lambda$ as a physical cutoff scale, one can think of it as a renormalisation scale \cite{Kondo:1992sq}.  For the model to be renormalisable it is clear that the fermion bilinear, $\psi\psi$, must have a large anomalous dimension, as the four-fermion operator is naively irrelevant.  This will indeed be the case.  Later on we will see that the anomalous dimension of $\psi\psi$ also determines the anomalous dimensions of the top partners.  Since large anomalous dimensions are required to couple the top partner to the elementary top quark, this approach is therefore much more appropriate.  A naive argument outlining its key features is as follows.

When $\bar{m}\ll\Lambda$ the VEV equation \eqref{eq:mmin} is well approximated by
\be
\frac{\bar{m}^2}{\Lambda^2}\ln\frac{\Lambda^2}{\bar{m}^2}\approx1-\frac{1}{\xi}
\ee
which, when $\Lambda$ is interpreted as a renormalisation scale, defines the RG dependence of $\xi$
\be
\beta(\xi)=\Lambda\pdiff{\xi}{\Lambda}\approx2\xi(1-\xi).
\ee
Hence there appears to be a UV fixed point (i.e.\ in the limit $\Lambda\to\infty$ with $\bar{m}$ finite) where $\xi=1$, corresponding to the critical point separating phases of broken and unbroken global symmetry.\footnote{There is also an IR fixed point at $\xi=0$, so the model cannot flow into the $\xi<0$ regime where the details of the vacuum become incalculable.}  For any RG flow starting near this critical point, $\xi$ naturally remains very close to one for a large range of energy with no need for tuning.

In addition, we can use the definition of the auxiliary scalar \eqref{eq:Mdef} to explicitly relate the fermion condensate, which is the physical order parameter for the spontaneous symmetry breaking, to the dynamically generated mass of the fermions, $\bar{m}$.  Assuming, for simplicity, that $\kappa_A>\kappa_B>0$, we have $\xi=\Lambda^2(\kappa_A+\kappa_B)/(4\pi^2)$ and
\be
\bar{m}=-\frac{4\pi^2\xi}{N_c\Lambda^2}\expect{\psi\psi}
\ee
(flavour indices have been suppressed; the full expression for $\expect{\psi\psi}$ has the same structure as eq.~\eqref{eq:M4vev} with the non-zero entries satisfying the above). Near the critical point, where $\xi\approx1$ is constant, this defines a running fermion mass in the broken phase
\be
\bar{m}(\Lambda)=\pfrac{\mu_0}{\Lambda}^2\bar{m}(\mu_0)\equiv Z_m\bar{m}(\mu_0)
\ee
where $\mu_0$ is a reference scale \cite{Yamawaki:1996vr}.  Defining an anomalous dimension for the mass then gives
\be
\gamma_m\equiv-\frac{\Lambda}{Z_m}\pdiff{Z_m}{\Lambda}=2
\ee
at the critical point, whereupon we find that the fermion bilinear $\psi\psi$ (i.e.\ the operator corresponding to the fermion mass term) must have scaling dimension $3-\gamma_m=1$.  The four-fermion operator therefore has scaling dimension 2, so the model appears to have become renormalisable.

Such a large anomalous dimension is a direct consequence of the existence of the UV fixed point.  In the absence of such a fixed point $\xi$ would continue to increase in the UV, scaling like $\Lambda^2$, because the four-fermion operator is naively irrelevant.  The dynamically generated fermion mass would then not run, so there would be no anomalous dimension (indeed, this is essentially what happens in the cutoff interpretation).  The presence of a fixed point curtails the running of $\xi$, so the dimensionality of the four-fermion couplings $\kappa_{A,B}$ are, in some sense, absorbed into $\psi\psi$ instead.

The reason that the argument presented so far is naive is that the mooted UV fixed point is actually trivial.  This is because fermion loops always generate kinetic and quartic terms for the auxiliary scalar, i.e.\ terms like $(\partial_\mu\psi\psi)^2$ and $(\psi\psi)^4$ in the original Lagrangian.  Since the fermion bilinear has scaling dimension 1 near the critical point, these terms are logarithmically divergent, just like in normal Yukawa theory (and as can be seen from the VEV equation \eqref{eq:mmin}).  However, unlike in normal Yukawa theory, there are no bare kinetic or quartic terms in the Lagrangian \eqref{eq:Leff} to absorb the divergences, so one is forced to rescale the auxiliary scalar to take care of them.  This means that the Yukawa coupling in eq.~\eqref{eq:Leff} scales like $1/\ln\Lambda$, so flows to zero in the UV\@.\footnote{The same logarithmic divergence can be seen in the full VEV equation, eq.~\eqref{eq:mmin}.}

A more thorough analysis, that considers the $Sp(2N_c)$ gauge coupling $\alpha$ as well as four-fermion operators by finding an approximate solution to the Schwinger-Dyson equations, suggests an alternative VEV equation.  This analysis is performed in Refs.~\cite{Bardeen:1985sm, Kondo:1988qd} for a $U(1)$ gauge theory with fixed gauge coupling, but their results can be extended to an $Sp(2N_c)$ gauge theory simply by adjusting the group theory factor to give
\be
\frac{2}{1-\omega^2}\pfrac{\bar{m}}{\Lambda}^{2\omega}=\frac{1}{\xi^*}-\frac{1}{\xi}
\ee
where
\begin{align}\label{eq:alphastar}
\xi^* & =\frac{1}{4}(1+\omega)^2 &
\omega & =\sqrt{1-\frac{\alpha}{\alpha^*}} &
\alpha^* & =\frac{\pi}{3C_2(\fund)}=\frac{2\pi}{3(2N_c+1)}.
\end{align}
It applies to the region in which $\alpha\le\alpha^*$, in both the broken and unbroken phases.  In addition, this analysis yields expressions for the beta function of $\xi$ and the anomalous dimension of the fermion mass \cite{Kondo:1992sq}
\begin{align}\label{eq:gammafull}
\beta(\xi) & =2\omega\xi\nbrack{1-\frac{\xi}{\xi^*}} &
\gamma_m(\xi) & =1-\omega\nbrack{1-\frac{2\xi}{\xi^*}}.
\end{align}
Perturbative running of the gauge coupling is accounted for by using the running value \cite{Miransky:1983vj, Higashijima:1983gx}
\be\label{eq:alphaRG}
\alpha(\Lambda)=\frac{2\pi}{b\ln(\Lambda/\Lambda_\alpha)}
\ee
where $\Lambda_\alpha$ is the strong coupling scale of the $Sp(2N_c)$ gauge group and $b$ is the beta function coefficient (evaluated in eq.~\eqref{eq:b65}).

\begin{figure}[!t]
\begin{center}
\includegraphics[width=0.4\textwidth]{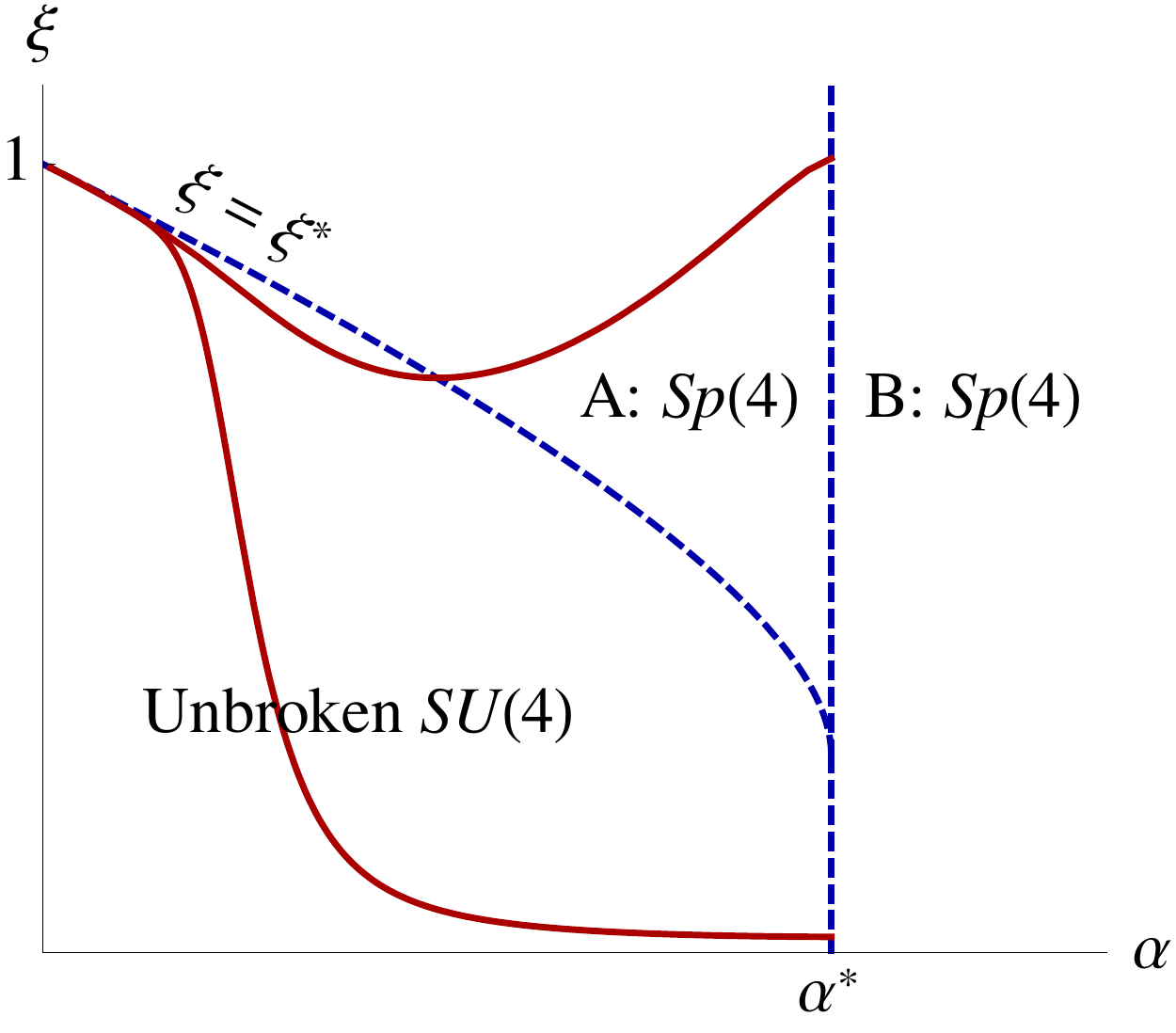}
\end{center}
\caption{The phase diagram in the $(\alpha,\xi)$ plane.  Dashed, blue lines denote contours of $\beta(\xi)=0$ and also separate regions of different symmetry breaking scenarios.  In region A $SU(4)$ is broken to $Sp(4)$ as a result of four-fermion operators and in region B the same symmetry breaking is induced by the large gauge coupling.  The solid, red lines show some example RG trajectories that realise the correct symmetry breaking pattern.  On the upper trajectory the symmetry breaking is driven by four-fermion operators and the anomalous dimension of the fermion bilinear is large.  On the lower trajectory the symmetry breaking is driven by the gauge coupling and the anomalous dimension of the fermion bilinear is small.\label{fig:phase65}}
\end{figure}

\begin{figure}[!t]
\begin{center}
\includegraphics[width=0.4\textwidth]{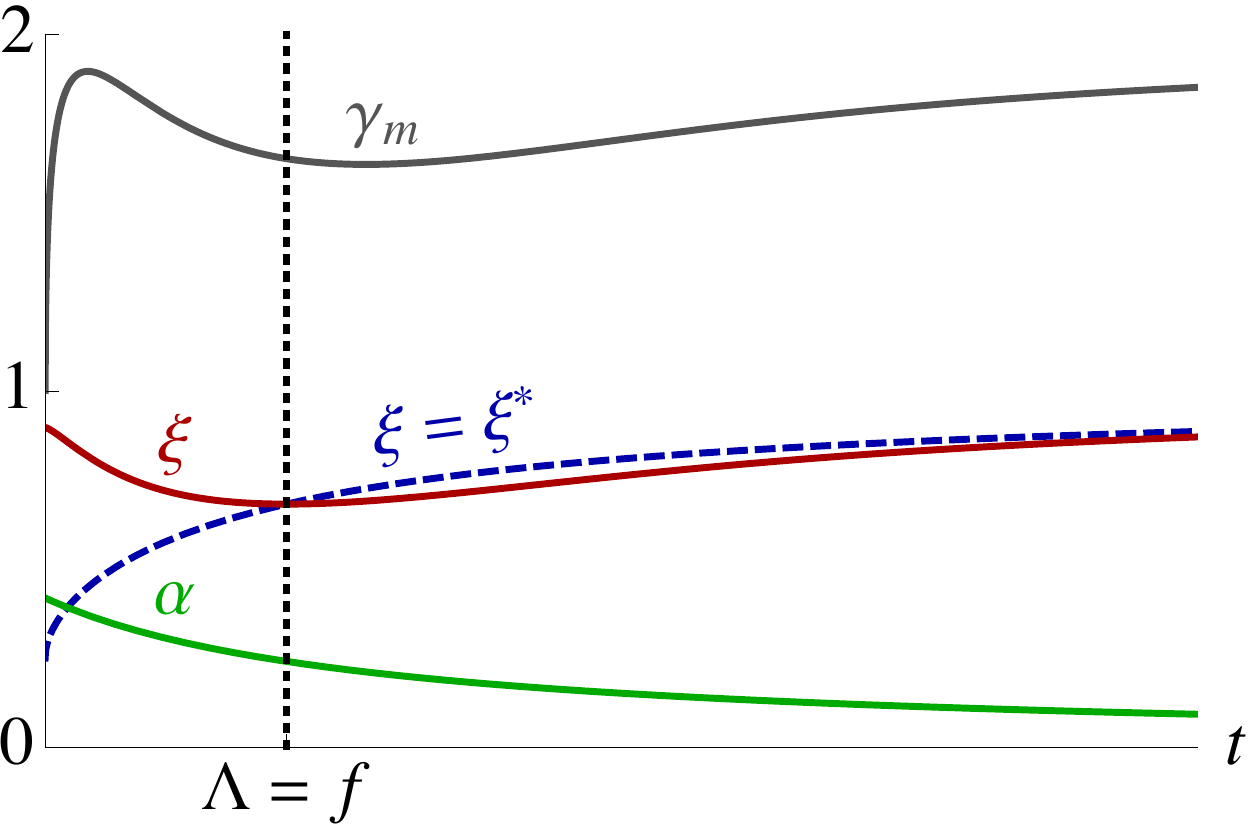}
\end{center}
\caption{The evolution of various quantities along the upper trajectory of figure \ref{fig:phase65} as a function of $t=\ln{(\Lambda/\Lambda_\alpha)}$.  The model crosses into the broken phase at the scale $\Lambda=f$.  The four-fermion coupling, $\xi$, remains close to the critical line, $\xi=\xi^*$, throughout the flow, and is much larger than the $Sp(2N_c)$ gauge coupling, $\alpha$.  Because of this, the anomalous dimension, $\gamma_m$, is very large.\label{fig:cplot}}
\end{figure}

The critical point at $\xi=1$ is therefore extended to a critical line in the $(\alpha,\xi)$ plane: $\xi=\xi^*$ for $\alpha<\alpha^*$, and $\alpha=\alpha^*$ for $\xi<\frac{1}{4}$.  The global symmetry is broken from $SU(4)$ to $Sp(4)$ either when $\xi>\xi^*$, or when $\alpha>\alpha^*$, leading to the phase diagram in figure \ref{fig:phase65}.  In the former case (the upper trajectory) the spontaneous symmetry breaking is driven primarily by four-fermion operators, and the anomalous dimension of the fermion bilinear can be large.  The evolution of various quantities along this trajectory is shown in figure \ref{fig:cplot}.  In the latter (the lower trajectory) it is being driven more by the gauge coupling, and the anomalous dimension of the fermion bilinear is small.  In fact, the latter case is essentially the same as treating $\Lambda$ as a UV cutoff scale.  Specifically, $\Lambda$ is then the scale at which the model's RG trajectory crosses the line $\alpha=\alpha^*$.

The gauge interactions decrease the anomalous dimension of the fermion mass from its previous value of 2 to lie somewhere in the range $1\le\gamma_m<2$, meaning that the scaling dimension of the fermion bilinear is now greater than 1 but less than or equal to 2.  Because of this, the logarithmic divergences of the kinetic and quartic terms for the auxiliary scalar are softened, the four-fermion operator still remains relevant, and it is possible for the model on the critical line line to be non-trivial and renormalisable~\cite{Kondo:1991yk, Harada:1994wy}.  To check whether this happens or not, one can consider the anomalous dimension near $\xi\approx\xi^*$ and for $\alpha\ll\alpha^*$
\be\label{eq:gamma}
\gamma_m\approx2-\frac{\alpha}{2\alpha^*}.
\ee
Thus the scaling of the fermion bilinear is decreased by
\be
(\psi\psi)_\Lambda=\exp\nbrack{\dimintlim{}{t}{\ln{\mu_0}}{\ln\Lambda}\gamma_m}(\psi\psi)_{\mu_0}=\pfrac{\Lambda}{\mu_0}^2\pfrac{\ln(\Lambda/\Lambda_\alpha)}{\ln(\mu_0/\Lambda_\alpha)}^{-\pi/(\alpha^*b)}(\psi\psi)_{\mu_0}
\ee
i.e.\ by a factor $(\ln\Lambda)^{-\pi/(\alpha^*b)}$.  This is sufficient to counter the logarithmic divergences of the kinetic and quartic terms if
\begin{align}
\frac{\pi}{\alpha^*b}>\frac{1}{2}\quad\implies\quad b<6C_2(\fund)
\end{align}
which is true for the model in question.  Adding extra matter, so as to produce the top partners, further slows the running of the gauge coupling so only improves the situation.

Another important consequence of the running gauge coupling can be seen in the example RG trajectories in figure \ref{fig:phase65}.  Since the gauge coupling invariably flows to larger values in the IR, it appears to be inevitable that all trajectories will, at some point, cross one of the two lines $\xi=\xi^*$ or $\alpha=\alpha^*$.  Thus the conclusion reached earlier, that spontaneous symmetry breaking only occurs if $\xi>1$, no longer applies.  Instead, symmetry breaking seems to happen on all possible RG trajectories.  In other words, confining gauge groups always want to induce fermion condensates in this kind of model.  This observation will have profound consequences when we consider more complicated models later.\footnote{Strictly speaking, this relies on using the perturbative expression \eqref{eq:alphaRG} for the running gauge coupling all the way to $\alpha=\alpha^*$.  However, we do not expect non-perturbative contributions to qualitatively change the behaviour, especially for large values of $N_c$ where $\alpha^*$ (defined in eq.~\eqref{eq:alphastar}) is small.}

Note that the argument just presented does not constitute a proof that a non-trivial UV description exists.  It simply suggests that the theory's existence is possible.  In particular, we had to appeal to approximations of non-perturbative physics (in this case an approximate solution to the Schwinger-Dyson equations) which have no guarantee of capturing the full behaviour of the UV description.  Rather, one is forced into considering these kinds of arguments if one wishes to use such a model to explain the Higgs sector.\footnote{Another possibility for a UV description is to run to a fixed point with large gauge coupling, as proposed in Ref.~\cite{Vecchi:2010jz}.  The fermion bilinear has dimension 2 at these fixed points, so they are less attractive for models that include top partners, as we shall soon see.}  This should be fully considered when assessing how attractive composite Higgs models are as a solution to the hierarchy problem.

\subsection{Top partners\label{sec:65c}}

The Higgs sector can, potentially, be described by the strong sector above.  However, the strong sector should also produce coloured, top partner bound states that can mix with the top quark.  Not only do these states facilitate strong coupling between the top and the composite Higgs, they are also required to explicitly break the NGB shift symmetry, allowing for a Higgs mass and potential.

Top partners can be constructed by adding a pair of fermions charged under $SU(3)$ colour.  A pair is necessary so that the new matter content is vector-like and does not render the Standard Model gauge group anomalous.  Consequently, the vectorial $SU(3)$ colour symmetry is enlarged to a chiral $SU(3)\times SU(3)$ symmetry, so the fermions must come with a Dirac mass term to explicitly break it back to its diagonal subgroup.  Otherwise we will end up introducing additional, unwanted NGBs to the spectrum.  Since the top quark carries hypercharge, the new fermions must be charged under an unbroken $U(1)$ symmetry as well.

It must also be decided how to construct fermionic bound states in this model.  There are no baryons in $Sp$ gauge theories as the group structure means that any baryons can be factorised into a combination of mesons.  Besides which, the number of colours is even so any baryon would be bosonic.  Hence the existence of a fermionic bound state requires fermions in a two (or more) index representation of the gauge group.

Putting everything together, we choose to add a vector-like pair of coloured fermions, $\chi$ and $\tilde{\chi}$, in the traceless (i.e. $\Omega_{ij}\chi_{ij}^f=\Omega_{ij}\tilde{\chi}_{ijf}=0$) antisymmetric representation of the $Sp(2N_c)$ gauge group.  The final matter content is then summarised in table \ref{tab:65c}.  The top partner candidates are
\begin{align}\label{eq:tpdef}
\Psi_1{}^{abf} & =(\psi^a\chi^f\psi^b) &
\Psi_2{}_{ab}^f & =(\bar{\psi}_a\chi^f\bar{\psi}_b) &
\Phi_{af}^b & =(\bar{\psi}_a\bar{\chi}_f\psi^b) \nonumber\\
\tilde{\Psi}_1{}^{abf} & =(\psi^a\tilde{\chi}_f\psi^b) &
\tilde{\Psi}_2{}_{ab}^f & =(\bar{\psi}_a\tilde{\chi}_f\bar{\psi}_b) &
\tilde{\Phi}_{af}^b & =(\bar{\psi}_a\bar{\tilde{\chi}}^f\psi^b)
\end{align}
with explicit index contractions $(\psi^a\chi^f\psi^b)=\psi_i^a\Omega_{ij}\chi_{jk}^f\Omega_{kl}\psi_l^b$ etc\@. and where $f$ is an $SU(3)$ colour index.  Note that the $\Psi$'s live in the antisymmetric ${\bf6}$ representation of $SU(4)$, which was shown to be an attractive choice of top partner representation in Ref.~\cite{Gripaios:2009pe}.  However, there are many other choices for the new fermions that allow for top partners in other $SU(4)$ representations.

\begin{table}[!t]
\begin{equation*}
\begin{array}{|c|c|cc|}\hline
\widerow & Sp(2N_c) & SU(4) & SU(3)_c\times U(1) \\\hline
\widerow \psi & \fund & {\bf4} & {\bf1}_0 \\
\widerow \chi & \asym & {\bf1} & {\bf3}_{+2/3} \\
\widerow \tilde{\chi} & \asym & {\bf1} & {\bf\bar{3}}_{-2/3} \\\hline\hline
\widerow M & {\bf1} & {\bf6} & {\bf1}_0 \\
\widerow S & {\bf1} & {\bf1} & {\bf1}_0 \\
\widerow R & {\bf1} & {\bf1} & {\bf8}_0 \\
\widerow P & {\bf1} & {\bf1} & {\bf6}_{+4/3} \\
\widerow \tilde{P} & {\bf1} & {\bf1} & {\bf\bar{6}}_{-4/3} \\\hline
\widerow \Psi_{1,2} & {\bf1} & {\bf6} & {\bf3}_{+2/3} \\
\widerow \Phi & {\bf1} & {\bf15}\oplus{\bf1} & {\bf\bar{3}}_{-2/3} \\
\widerow \tilde{\Psi}_{1,2} & {\bf1} & {\bf6} & {\bf\bar{3}}_{-2/3} \\
\widerow \tilde{\Phi} & {\bf1} & {\bf15}\oplus{\bf1} & {\bf3}_{+2/3} \\\hline
\end{array}
\end{equation*}
\caption{A model that realises the spontaneous symmetry breaking pattern $SU(4)\to Sp(4)$ and produces coloured, fermionic bound states.  $\psi$, $\chi$ and $\tilde{\chi}$ are elementary, two component Weyl fermions, $M$, $S$, $R$, and the $P$'s are complex scalar mesons, and the $\Psi$'s and $\Phi$'s are two component Weyl fermion mesons.\label{tab:65c}}
\end{table}

The new matter comes in pairs as far as $Sp(2N_c)$ is concerned, so the sum of the Dynkin indices remains even and no Witten anomaly is introduced.  In addition, the $Sp(2N_c)$ gauge group remains asymptotically free provided $N_c<19$, as one then has
\be
b=\frac{11}{3}(2N_c+2)-\frac{2}{3}(4+6(2N_c-2))=\frac{2}{3}(19-N_c)>0
\ee
where $T\nbrack{\asym}=2N_c-2$ has been used.

The interaction Lagrangian for the extended model contains additional scalar-scalar terms
\begin{align}
\cL_{\rm int}\supset{} & \sbrack{m_\chi\chi^f\tilde{\chi}_f+\mbox{h.c.}}+\frac{\kappa_S}{2N_c}(\chi^f\tilde{\chi}_f)(\bar{\chi}_g\bar{\tilde{\chi}}^g)+\frac{\kappa_R}{2N_c}(\chi^f\tilde{\chi}_g)_{{\rm tr}=0}(\bar{\chi}_f\bar{\tilde{\chi}}^g)_{{\rm tr}=0}+{} \nonumber\\
& \frac{\kappa_P}{2N_c}(\chi^f\chi^g)(\bar{\chi}_f\bar{\chi}_g)+\frac{\kappa_{\tilde{P}}}{2N_c}(\tilde{\chi}_f\tilde{\chi}_g)(\bar{\tilde{\chi}}^f\bar{\tilde{\chi}}^g).
\end{align}
Following the steps taken in the previous section then leads to
\begin{align}
\cL_{\rm int}\supset & -\sbrack{S^*(\chi^f\tilde{\chi}_f)+R_f^{g*}(\chi^f\tilde{\chi}_g)_{{\rm tr}=0}+P^*_{fg}(\chi^f\chi^g)+\tilde{P}^{fg*}(\tilde{\chi}_f\tilde{\chi}_g)+\mbox{h.c.}}-{} \nonumber\\
& {}-\frac{2N_c}{\kappa_S}|S+m_\chi^*|^2-\frac{2N_c}{\kappa_R}R^{g*}_fR^f_g-\frac{2N_c}{\kappa_P}P^*_{fg}P^{fg}-\frac{2N_c}{\kappa_{\tilde{P}}}\tilde{P}^{fg*}\tilde{P}_{fg}
\end{align}
where the coloured, auxiliary scalars
\begin{align}
S & =-\frac{\kappa_S}{2N_c}(\chi^f\tilde{\chi}_f)-m_\chi^* & R^f_g & =-\frac{\kappa_R}{2N_c}(\chi^f\tilde{\chi}_g)_{{\rm tr}=0} \nonumber\\
P^{fg} & =-\frac{\kappa_P}{2N_c}(\chi^f\chi^g) & \tilde{P}_{fg} & =-\frac{\kappa_{\tilde{P}}}{2N_c}(\tilde{\chi}_f\tilde{\chi}_g)
\end{align}
have been introduced.  These describe the VEVs of all coloured, scalar meson fermion bilinears in the model.  $Sp(2N_c)$ indices are again contracted with $\Omega$ (see eq.~\eqref{eq:Omega}) to yield auxiliary scalars transforming as in table \ref{tab:65c}.  It is important that $M$ and $S$ are the only scalars allowed to get a non-zero VEV, otherwise $SU(3)$ colour will be broken.

Deriving the one-loop effective potential for the coloured sector is more involved than before as the mass matrix for $\chi$ and $\tilde{\chi}$ is more complicated.  However, it is clear that the desired VEV pattern is easily attainable when the $Sp(2N_c)$ gauge coupling is neglected.  All fermion mass terms are proportional to auxiliary scalar VEVs, so the one-loop contribution to the potential does not contain any linear terms for the scalars.  As long as the $\kappa$'s in the coloured sector are all small enough, the tree level mass-squared term for each scalar (proportional to $2N_c/\kappa$) will then dominate the one-loop contribution (proportional to $N_c\Lambda^2$).  Hence all coloured scalars will be stabilised at the origin, except possibly $S$ as there is a linear term for it in the potential.

If $\Lambda$ is interpreted as a physical UV cutoff, this analysis suffices to be confident that the correct symmetry breaking pattern is observed.  The problem then is that all anomalous dimensions remain too small, and the four-fermion operators mixing top partners with elementary tops will be irrelevant.

\begin{figure}[!t]
\begin{center}
\includegraphics[width=0.4\textwidth]{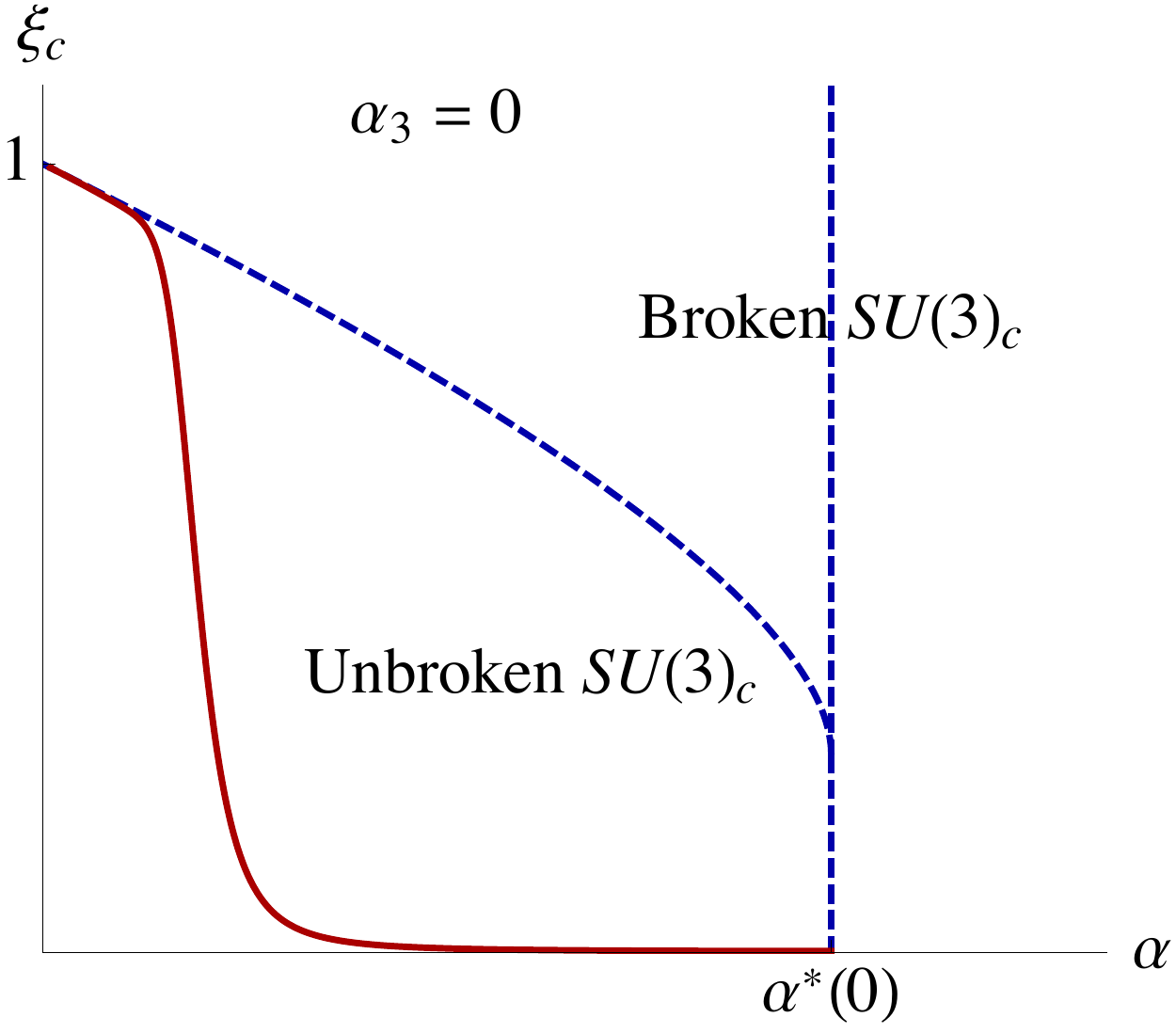}\hspace{10mm}
\includegraphics[width=0.4\textwidth]{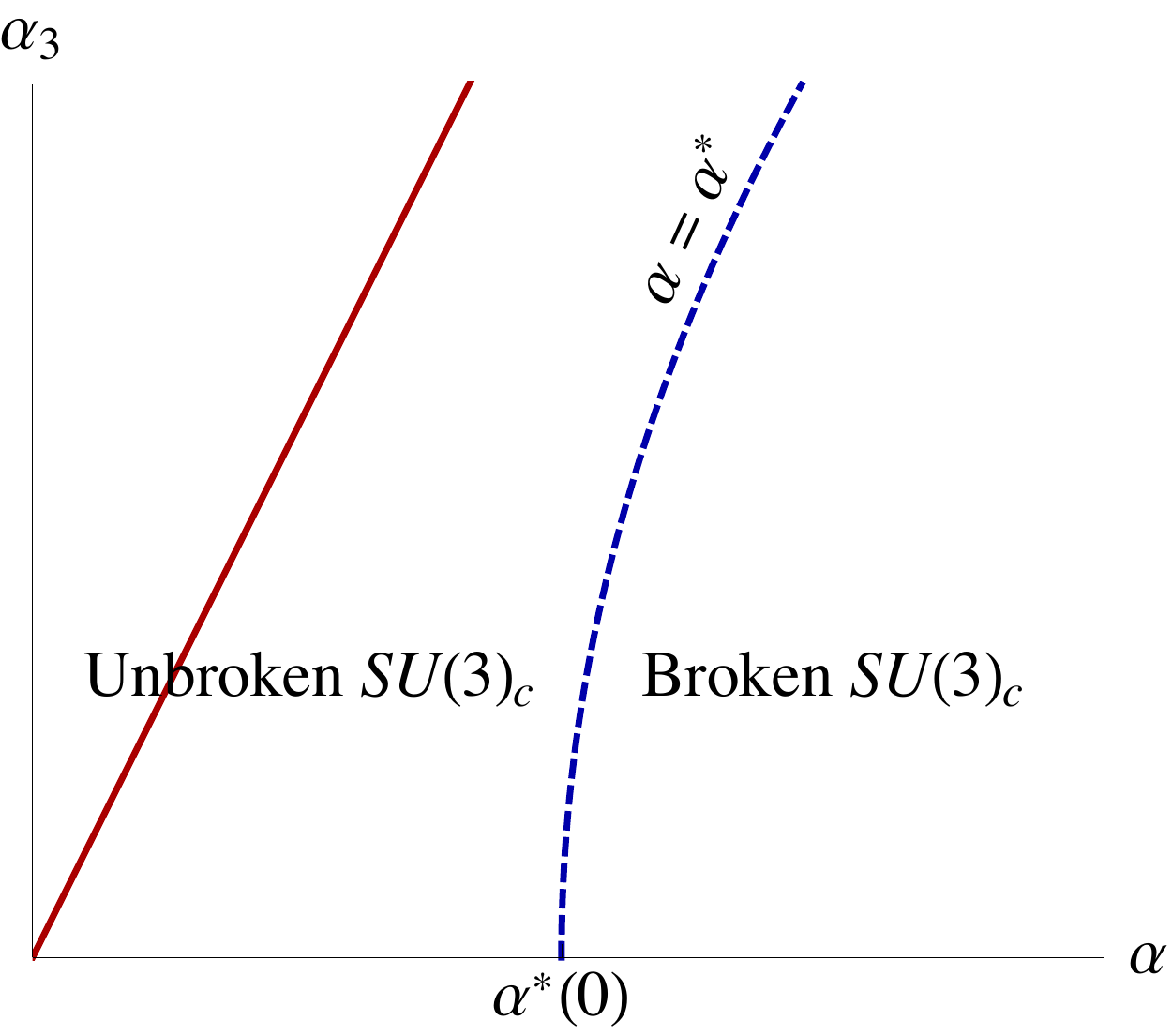}
\end{center}
\caption{A possibility for the phase diagram in the coloured sector.  The coupling $\xi_c$ controls the strength of four-fermion interactions in this sector and is defined by an expression similar to eq.~\eqref{eq:mmin}. Dashed, blue lines separate different symmetry breaking scenarios and the solid, red lines show some example RG trajectories that could realise the correct symmetry breaking pattern.\label{fig:phase65c}}
\end{figure}

An analysis that includes the effects of running couplings as in section \ref{sec:LA} could change this conclusion, as it did in the colour neutral sector.  To be sure, we would need expressions for the two-point functions $\expect{\chi\chi}$, $\expect{\tilde{\chi}\tilde{\chi}}$ and $\expect{\chi\tilde{\chi}}$.  Since $\chi$ and $\tilde{\chi}$ are in two index representations of the gauge group, we cannot simply recycle the results of Ref.~\cite{Kondo:1992sq} with a different group theory factor, as we did previously for the two-point function $\expect{\psi\psi}$.  Nonetheless, assuming that the phase diagram corresponding to each of the four-fermion operators in the coloured sector is qualitatively similar to figure \ref{fig:phase65}, flows such as those on the left of figure \ref{fig:phase65c} would be obtained.  Even if the $\kappa$'s are small enough at some point on the RG trajectory for $SU(3)$ colour to remain unbroken, the confining $Sp(2N_c)$ gauge group induces fermion condensates and pushes the model into the broken phase.

However, this phase diagram neglects the effects of a non-zero QCD gauge coupling, $\alpha_3$.  Because coloured fermion condensates break $SU(3)$ colour, they also give a mass (proportional to $\sqrt{\alpha_3}$) to the QCD gluons.  This means that the gluons provide an additional contribution to the one-loop effective potential.  In particular, this contribution acts to stabilise the coloured, auxiliary scalar potential around the origin, with the effect being larger for larger values of $\alpha_3$.  Hence, when the QCD gauge coupling is taken into account, the previously fixed constant $\alpha^*$ should be replaced by an increasing function of $\alpha_3$.  As the model flows into the IR we know that the QCD gauge coupling (ergo the stabilising effect of the gluons) increases, so one may hope to avoid a phase transition after all.  The situation would have to be something like that illustrated on the right of figure \ref{fig:phase65c}.  It is only viable if the ratio $\alpha/\alpha_3$ (the inverse slope of the solid, red line on the right of figure \ref{fig:phase65c}) is smaller than ${\rm d}\alpha^*/{\rm d}\alpha_3$ (the inverse slope of the dashed, blue line) as all couplings flow into the IR\@.

Breaking $SU(3)$ colour may also be avoidable if the bare mass term for the coloured fermions, $m_\chi$, is much larger than the scale at which $\alpha=\alpha^*$.  The coloured fermions could then be integrated out before the RG trajectory is forced to cross the critical line, leaving nothing to form a coloured, symmetry breaking condensate out of.  However, this approach may be at odds with the requirement that the low-energy theory contains light top partners.  If all coloured fermions are integrated out too soon, there is nothing to form light top partners out of either (c.f.\ the relatively large mass of bottom quark bound states in QCD).  One would be forced to perform a balancing act, choosing $f\gg m_\chi\gg\Lambda_{\alpha=\alpha^*}$, such that top partners can be lighter than the spontaneous symmetry breaking scale, $f$, but a coloured fermion condensate is still prevented from developing.  While this solution seems somewhat ad hoc, it is clear from the upper trajectory in figure \ref{fig:phase65} that $\alpha$ can be much less than $\alpha^*$ when the RG trajectory in the colour neutral sector crosses the critical line (i.e.\ $f\gg\Lambda_{\alpha=\alpha^*}$).  Indeed, if the top partners are to have a large anomalous dimension, eq.~\eqref{eq:gamma} suggests that this kind of trajectory, with $\xi\approx\xi^*$, is preferred.  It is then only $m_\chi$ that is subject to an additional constraint.

Even then, it may not be possible to avoid a coloured condensate entirely.  The effect of the bare mass term may just be to suppress the condensate by powers of $\Lambda_{\alpha=\alpha^*}/m_\chi$.  Given the extremely strict limits on the mass of the gluon, a very large suppression would be required, restoring unwanted tuning in the model.  To truly see how viable it is to avoid coloured condensates by using a bare mass term, a full RG analysis for massive fermions in the antisymmetric representation would again have to be performed.

\subsubsection{Anomalous dimension of the top partner\label{sec:ADtop}}

The elementary top is typically coupled to the strong sector via mixing terms of the form
\be
\cL\supset\lambda_L\Psi_Lt_L+\lambda_R\Psi_Rt_R
\ee
where $t_{L,R}$ are (two component) fermions representing the left and right handed top quarks.  The top partners, $\Psi_{L,R}$ (each one of the options from eq.~\eqref{eq:tpdef}), are themselves three-fermion operators thus, in addition to the four-fermion operators already used to drive the spontaneous global symmetry breaking, the mixing terms are four-fermion operators too.  If the scaling dimension of either top partner is too large, these operators become irrelevant, leading to a too small top quark Yukawa coupling and a failure to trigger electroweak symmetry breaking.  Specifically, the top mass scales like \cite{Contino:2010rs}
\be
m_t\propto\lambda_L\lambda_R
\ee
where the values of the $\lambda$'s at the weak scale are determined by the dimension of the corresponding top partner
\be
\lambda_{L,R}\propto\pfrac{\mbox{TeV}}{\Lambda}^{\dim{\Psi_{L,R}}-\frac{5}{2}}.
\ee
If $\dim{\Psi}\gtrsim\frac{5}{2}$, the $\lambda$'s are power suppressed by the RG flow and the top mass may be too small.

A full quantitative description of the top partner scaling dimension is beyond the scope of this paper, but one can follow the diquark model of baryons \cite{Lichtenberg:1968zz} to motivate a plausibility argument.  Generally, there are three separate contributions to the anomalous dimension of the top partner; one from four-fermion operators in the colour neutral sector (e.g.\ $(\psi\psi)(\psi\psi)$), one from four-fermion operators in the coloured sector (e.g.\ $(\chi\chi)(\chi\chi)$), and one from $Sp(2N_c)$ gauge interactions.

When four-fermion operators in the colour neutral sector drive spontaneous symmetry breaking, and $SU(3)$ colour remains unbroken, the RG trajectory of the model should look like the one represented by the upper line in figure \ref{fig:phase65} in the colour neutral sector and the one on the left of figure \ref{fig:phase65c} in the coloured sector.  Hence four-fermion interactions in the colour neutral sector are much stronger than $Sp(2N_c)$ gauge interactions ($\xi\gg\sqrt{\alpha}$ as in figure \ref{fig:cplot}), and these are much stronger than four-fermion interactions in the coloured sector ($\sqrt{\alpha}\gg\xi_c$).  Consequently, we can consider three fermion states of the form $\psi\chi\psi$ to be composed of a tightly bound diquark, $\psi\psi$, combined with an additional quark, $\chi$, that is much more weakly bound.  Their scaling dimension is then approximated by adding the scaling dimension of the diquark to the canonical scaling dimension ($3/2$) of $\chi$.

The approximate solution to the Schwinger-Dyson equations turns out to be the same for a diquark fermion bilinear as it was for the scalar meson fermion bilinear, despite the different index structure \cite{Ball:1990gf}.  The scaling dimensions of the two operators are therefore equal, meaning that the diquark also has scaling dimension $3-\gamma_m$, with $1\le\gamma_m<2$ given by eq.~\eqref{eq:gammafull} or, in the regime of interest, eq.~\eqref{eq:gamma}.  This is enough for the scaling dimension of the top partner to be arbitrarily close to, albeit slightly above $5/2$, so that the $\Psi t$ mixing operator is almost relevant.

Even though the mixing operator is not quite relevant, a sufficiently large top quark Yukawa coupling can still be accommodated.  One can numerically solve the RG equations for the matter content of section \ref{sec:65c} with $N_c=2$, a spontaneous symmetry breaking scale of 10 TeV, and a strong coupling scale of $\Lambda_c=1$ TeV for the $Sp(2N_c)$ gauge coupling.  We find a suppression of about 10\% for $\lambda_{L,R}$ when running down from the Planck scale, so an initial coupling of order $10$ yields a top quark Yukawa coupling of order one. This is consistent with the non-perturbative nature of the UV description.  It is also conceivable that the effects ignored so far (four-fermion interactions in the colour neutral sector and $Sp(2N_c)$ gauge interactions), both of which naively decrease the scaling dimension of the top partner, will push it below the critical value of $5/2$.

Finally, we point out that the third fermion is glued in place by $Sp(2N_c)$ gauge interactions, which then control the residual strong interaction between the top partners and the composite Higgs, analogously to the large meson-baryon couplings in QCD \cite{Ball:1990gf}.  Since gauge interactions are much weaker than four-fermion interactions at the confinement scale (see figure \ref{fig:cplot}), the mass of the top partners is expected to be lower than that of colour neutral resonances bound only by four-fermion interactions.  This is helpful for reproducing the observed Higgs mass \cite{Matsedonskyi:2012ym, Marzocca:2012zn, Pomarol:2012qf, Panico:2012uw, Pappadopulo:2013vca, Barnard:2013hka}.

\subsection{Observables in the UV description\label{sec:UVobs}}

The UV description of the $SO(6)/SO(5)$ model, which contains the $Sp(2N_c)$ fermions $\psi$, $\chi$, and $\tilde\chi$ charged under the SM gauge group, can be used to compute a number of observables.  First, we can consider the effects on the running of the SM gauge couplings. When $N_c > 2$ we find that a Landau pole in the SM gauge couplings develops below the GUT scale ($\simeq 10^{16}$ GeV) due to the multiplicity of fermions from the $Sp(2N_c)$ gauge group.  However, this is no longer true for $N_c=2$, leading to a unique choice for which the overall UV description remains well defined (the case $N_c=1$ has no antisymmetric fermion representations).

In the $N_c=2$ case, even though the gauge couplings run without Landau poles, there is no improvement in their unification compared to the SM\@.  However the $\psi$ fermions can help to improve the unification in the SM (although not at the same level as in the MSSM), provided that $\chi$ and $\tilde\chi$ do not contribute to the relative running.  This therefore depends on the ad-hoc assumption of introducing further spectator fermions to fill out complete $SU(5)$ multiplets, or assuming that the strong sector couples to the SM without top partners~\cite{Galloway:2010bp}.  If spectator fermions are included, for example by embedding $\chi,\tilde \chi$ into the ${\bf10}+{\bf\overline{10}}$ representation of $SU(5)$, then an approximate unification occurs at a reduced GUT scale around $10^{12}$ GeV, but proton decay problems would need to be addressed.

The UV description also allows for explicit calculation of the contributions from the strong sector to other observables. These include the oblique electroweak precision observables, $S$ and $T$, various flavour observables, and nontrivial relationships between the Higgs and top partner masses.  Also of significance is the contribution of the strong sector to the Higgs couplings beyond the Higgs boson low energy theorems.  In the present context, all of this depends on the coupling of the top partner with the composite Higgs.  This is proportional to the $Sp(2N_c)$ gauge coupling $\alpha$ and can be estimated in a similar way to the meson-baryon couplings in QCD\@.  These calculations are beyond the scope of this paper and will be left for future work.

\section{Other composite Higgs models\label{sec:other}}

Several important lessons can be learnt from the example model above, and they provide some clues about how likely it might be to UV complete other composite Higgs models.  Perhaps the two most important lessons are that one must think carefully about what sort of scalar mesons can be constructed from an underlying theory of fermions, and that confining gauge groups induce fermion condensates wherever possible.  To see how these lessons apply, it is instructive to consider the minimal composite Higgs model, in which an $SO(5)$ global symmetry is spontaneously broken to $SO(4)$ by the VEV of a scalar in the fundamental representation.

An immediate complication comes from the fact that the original global symmetry, $SO(5)$, is not simply an exchange symmetry between massless fermions, which are complex objects.  To realise it, a genuine exchange symmetry, such as the $SU(5)$ symmetry acting on five identical fermions, must be explicitly broken to $SO(5)$ by operators in the Lagrangian.  These could be fermion bilinears, whereupon the fermions transforming under $SO(5)$ end up vector-like and massive.  Or they could be four-fermion operators, whereupon the fermions can remain massless but one is forced into arguments such as those in section \ref{sec:LA}; the four-fermion operator cannot be considered an effective operator if it explicitly breaks a symmetry, as the question of UV description is then simply shifted into another sector.  This problem persists in any model in which the original global symmetry is not an $SU$ group.

An additional (and more serious) complication is the existence of more than one scalar meson in any potential UV description.  If a scalar meson in the fundamental representation of $SO(5)$ can be constructed out of two fermions, there will always be another scalar meson in a different representation.  For example, if the UV description contains a fermion, $\psi$, in the fundamental representation of $SO(5)$, and a fermion, $\lambda$, in the singlet representation, there are three scalar mesons: $\psi\lambda$ in the fundamental representation, $\lambda\lambda$ in the singlet representation, and $\psi\psi$ in some other representation.  Only $\psi\lambda$ and $\lambda\lambda$ can form condensates, otherwise the symmetry breaking pattern will be ruined.  However, we saw in section \ref{sec:LA} that confining gauge groups generically like to induce all fermion condensates possible. 

This problem was already encountered when we tried to add top partners to the $SU(4)\to Sp(4)$ model in section \ref{sec:65c}.\footnote{The problem is less acute here as the unwanted condensate does not produce a gluon mass.  A small $\psi\psi$ condensate can probably be tolerated, provided the associated NGBs are weakly coupled to the rest of the theory.  A bare mass term for $\psi$ is therefore a more reasonable option this time around.}  To get around it there we could appeal to the stabilising effect of QCD gauge interactions, as the extra scalar mesons were charged under $SU(3)$ colour.  This will not work for scalar mesons such as $\psi\psi$, as they are not charged under any gauge group.  Another possibility, also discussed in section \ref{sec:65c}, is to include a bare mass term for $\psi$.  As we saw previously, this necessitates a balancing act between scales; $f\gg m_\psi\gg\Lambda_{\alpha=\alpha^*}$, where $f$ is the spontaneous symmetry breaking scale, $m_\psi$ is the $\psi$ mass, and $\Lambda_{\alpha=\alpha^*}$ is the scale at which the confining gauge group alone induces a fermion condensate.  This allows $\psi$'s to remain in play long enough to form a $\psi\lambda$ condensate and spontaneously break $SO(5)$ to $SO(4)$, but not to form a $\psi\psi$ condensate.

With these two issues in mind, a possible UV description of the minimal model is outlined in table \ref{tab:54}.  The interaction Lagrangian for this model would have to be something like
\begin{align}\label{eq:L54}
\cL_{\rm int}={} & \frac{\kappa_{MA}}{N_c}(\psi_a\psi_b)(\bar{\psi}_a\bar{\psi}_b)+\frac{\kappa_{MB}}{N_c}(\psi_a\psi_a)(\bar{\psi}_b\bar{\psi}_b)+\frac{\kappa_{MC}}{2N_c}\sbrack{(\psi_a\psi_b)(\psi_a\psi_b)+\mbox{h.c.}}+{} \nonumber\\
& \frac{\kappa_{FA}}{N_c}(\psi_a\lambda)(\bar{\psi}_a\bar{\lambda})+\frac{\kappa_{FB}}{2N_c}\sbrack{(\psi_a\lambda)(\psi_a\lambda)+\mbox{h.c.}}+{} \nonumber\\
& \frac{\kappa_{SA}}{N_c}(\lambda\lambda)(\bar{\lambda}\bar{\lambda})+\frac{\kappa_{SB}}{2N_c}\sbrack{(\lambda\lambda)(\lambda\lambda)+\mbox{h.c.}}+m_\psi(\psi_a\psi_a)+m_\lambda(\lambda\lambda)
\end{align}
where $a,b$ are $SO(5)$ indices, the $\kappa$'s are real coupling constants, and the masses for the fermions satisfy $f\gg m_\psi\gg\Lambda_{\alpha=\alpha^*}$ and $f\gg m_\lambda$.  The mass term for $\psi$ means that the fermions are now real, Majorana fermions, explicitly breaking the original $SU(5)$ fermion exchange symmetry to $SO(5)$, the equivalent exchange symmetry acting on real objects.  A full analysis of the model would proceed much as the analysis presented in section \ref{sec:65}, although with a more complicated one-loop effective potential and a larger coloured sector (any $\psi$ could be replaced with a $\lambda$ in eq.~\eqref{eq:tpdef}).  The top partners will also be in several different representations.  While it seems possible that everything can be arranged to work out, the additional complexity, larger number of free parameters, and increased reliance on approximations of non-perturbative physics make the model much less appealing than the $SU(4)\to Sp(4)$ model of section \ref{sec:65}.

\begin{table}[!t]
\begin{equation*}
\begin{array}{|c|c|c|}\hline
\widerow & SO(N_c) & SO(5) \\\hline
\widerow \psi & \fund & {\bf5} \\
\widerow \lambda & \fund & {\bf1} \\\hline\hline
\widerow M & {\bf1} & {\bf10} \\
\widerow F & {\bf1} & {\bf5} \\
\widerow S & {\bf1} & {\bf1} \\\hline
\end{array}
\end{equation*}
\caption{A model that may realise the spontaneous symmetry breaking pattern $SO(5)\to SO(4)$.  $\psi$ and $\lambda$ are two component Weyl fermions, and $M$, $F$ and $S$ are complex scalar mesons.\label{tab:54}}
\end{table}

The issues encountered in the minimal model will also be present for certain other choices of the coset space $G/H$, suggesting that some models have better prospects for UV description than others.  Dealing with initial global symmetries that are not $SU$ fermion exchange symmetries is not too serious a concern, although it will always increase model complexity by requiring explicit symmetry breaking operators in the interaction Lagrangian.  A bigger problem is to limit the number of scalar mesons so as to preserve the symmetry breaking pattern.  Ideally, there should be only one representation of scalar meson in the spectrum to avoid this problem altogether.

Models in which the spontaneous symmetry breaking is due to a scalar meson in the symmetric or antisymmetric representation of $G$ are very amenable from this point of view.  Such mesons are naturally formed in $SO$ and $Sp$ gauge theories respectively, where they are the only mesons in the spectrum.  Spontaneous symmetry breaking due to a meson in the adjoint representation is another attractive option.  These are formed in $SU$ gauge theories with vector-like matter content, and again are the only mesons in the spectrum.  Even so, there are other complications that may arise.  As an example, consider the model with symmetry breaking pattern $SU(5)\to SU(4)$.  Although this symmetry breaking can be due to a scalar meson in the symmetric representation, it is unclear how the required VEV could be generated; the potential for the meson's eigenvalues would probably need to have two degenerate minima, one at zero and one elsewhere, suggesting tuning in the model.

\section{Summary\label{sec:sum}}

The idea of a composite Higgs boson remains a interesting possibility, especially if the Higgs is identified as the NGB of a spontaneously broken global symmetry and the top is partially composite.  Effective, low-energy models have been constructed, primarily based on the AdS/CFT correspondence, and have been quite successful in describing low-energy phenomenology.  However a complete UV description remains unexplored, except for supersymmetric UV descriptions based on Seiberg duality.

In an attempt to address this issue in a nonsupersymmetric way we have considered a simple, nonminimal composite Higgs model based on the coset $SO(6)/SO(5)$.  The underlying strong dynamics is described by an $Sp(2N_c)$ gauge group with four flavours of Weyl fermion, as well as a pair of vector-like fermions in the antisymmetric representation which are also triplets of QCD\@.  Four-fermion operators are added to this gauge theory, like in the gauged NJL model, and trigger the spontaneous breaking of the $SU(4)\simeq SO(6)$ global symmetry to $Sp(4)\simeq SO(5)$.  A particularly interesting consequence of this breaking is that it occurs in the presence of a large anomalous dimension for the constituent fermion bilinear, meaning that the composite top partner operator also has a large anomalous dimension, therefore leading to a sizeable top quark Yukawa coupling.  This provides a simple UV description of the dynamics responsible for all features of the $SO(6)/SO(5)$ composite Higgs model.

Our methods can also be applied to other coset groups, in particular the minimal $SO(5)/SO(4)$ model.  The details of the corresponding NJL model requires a more comprehensive analysis.  However, qualitative arguments suggest that the underlying features of the strong dynamics are similar to the $SO(6)/SO(5)$ model.  It would be interesting to explore these models further.  In any case, a composite Higgs remains an intriguing option and four-fermion operators provide one way of understanding the underlying strong dynamics.

\subsubsection*{Acknowledgements}

We thank Steven Abel, Carlos Savoy, Misha Shifman, and Arkady Vainshtein for useful discussions. This work was supported by the Australian Research Council. TG was also supported in part by the DOE grant DE-FG02-94ER-40823. JB and TG thank the Galileo Galilei Institute for Theoretical Physics for hospitality and the INFN for partial support during the completion of this work.

\appendix

\section{Minimising the $SU(4)\to Sp(4)$ potential\label{app:pot}}

All minima of eq.~\eqref{eq:VSp1} satisfy
\begin{align}
\pdiff{V}{\bar{m}_1}={} & \frac{2N_c\kappa_A}{\kappa_A^2-\kappa_B^2}\bar{m}_1-\norm{\frac{2N_c\kappa_B}{\kappa_A^2-\kappa_B^2}}\bar{m}_2-\frac{N_c}{2\pi^2}\bar{m}_1\sbrack{\Lambda^2-\bar{m}_1^2\ln\pfrac{\Lambda^2+\bar{m}_1^2}{\bar{m}_1^2}}=0 \nonumber\\
\pdiff{V}{\bar{m}_2}={} & \frac{2N_c\kappa_A}{\kappa_A^2-\kappa_B^2}\bar{m}_2-\norm{\frac{2N_c\kappa_B}{\kappa_A^2-\kappa_B^2}}\bar{m}_1-\frac{N_c}{2\pi^2}\bar{m}_2\sbrack{\Lambda^2-\bar{m}_2^2\ln\pfrac{\Lambda^2+\bar{m}_2^2}{\bar{m}_2^2}}=0
\end{align}
which can be written as
\begin{align}\label{eq:fsol}
f(\bar{m}_1) & =\bar{m}_2 & f(\bar{m}_2) & =\bar{m}_1
\end{align}
for $f(m)$ of the form
\be
f(m)=am\nbrack{m^2\ln\pfrac{m^2+\Lambda^2}{m^2}-b}
\ee
where $a$ and $b$ are real constants, $a>0$, and $\bar{m}_1$ and $\bar{m}_2$ are both positive.  Positivity of $\bar{m}_1$ and $\bar{m}_2$ restrict all non-zero solutions of eqs.~\eqref{eq:fsol} to lie in the domain in which $f(m)$ is strictly positive, i.e.\
\be
m^2\ln\pfrac{m^2+\Lambda^2}{m^2}>b.
\ee
The derivative of $f(m)$ is
\be
f^\prime(m)=a\nbrack{3m^2\ln\pfrac{m^2+\Lambda^2}{m^2}-b-\frac{2m^2\Lambda^2}{m^2+\Lambda^2}}.
\ee
When $f(m)$ is strictly positive, one therefore has
\be
f^\prime(m)>2am^2\nbrack{\ln\pfrac{m^2+\Lambda^2}{m^2}-\frac{\Lambda^2}{m^2+\Lambda^2}}\ge0
\ee
with equality only at $m=0,\infty$.  We thus conclude that $f(m)$ is strictly increasing over the domain of $m$ for which $f(m)$ is strictly positive, hence is a function (in the formal mathematical sense) over this domain, so has a unique inverse.  This means that the two equations reduce to
\be
f(\bar{m}_1)=\bar{m}_2=f^{-1}(\bar{m}_1)
\ee
i.e.\ they describe a function intersecting its inverse.  Solutions to this problem only exist for $f(m)=m$ due to the symmetry of the problem, which implies $\bar{m}_1=\bar{m}_2$ in all solutions to eqs.~\eqref{eq:fsol}.

\bibliographystyle{JHEP-2}
\bibliography{UVNGB}
\end{document}